\documentclass[aoas,secthm,seceqn,nameyear,noautosecdot,dvips]{arximspdf}
\usepackage{graphicx}

\doi{10.1214/10-AOAS399}
\volume{5}
\issue{2A}
\pubyear{2011}
\firstpage{943}
\lastpage{968}

\makeatletter
\let\epsilon\varepsilon
\newcommand{\eps}{\epsilon}
\newcommand{\hatm}{\hat{m}_0}
\newcommand{\hatmone}{\hat{m}_0^{(\not1)}}
\newcommand{\galm}{\tilde{m}_0}
\newcommand{\galmone}{\tilde{m}_0^{(\not1)}}
\newcommand{\vecp}{\vec{p}}

\newproclaim{definition}[thm]{Definition}
\newtheorem{claim}[thm]{Claim}
\newtheorem{corollary}[thm]{Corollary}
\newproclaim{example}[thm]{Example}

\makeatother

\begin{document}
\begin{frontmatter}

\title{FDR control with adaptive procedures and FDR~monotonicity\thanksref{TT}}
\runtitle{FDR control with adaptive procedures and FDR monotonicity}
\thankstext{TT}{Supported by the Ridgefield Foundation,
the Mitchell Foundation and EC FP6 Contract No. 502983.}

\begin{aug}
\author[A]{\fnms{Amit} \snm{Zeisel}\thanksref{t1}\ead[label=e1]{amit.zeisel@weizmann.ac.il}},
\author[B]{\fnms{Or} \snm{Zuk}\thanksref{t1}\ead[label=e2]{orzuk@broadinstitute.org}}
\and
\author[C]{\fnms{Eytan} \snm{Domany}\corref{}\thanksref{t3}\ead[label=e3]{eytan.domany@weizmann.ac.il}
\ead[label=u1,url]{http://www.foo.com}}
\thankstext{t1}{Equal contribution.}
\thankstext{t3}{Corresponding author.}
\runauthor{A. Zeisel, O. Zuk and E. Domany}
\affiliation{Weizmann Institute, Broad
Institute and Weizmann Institute}
\address[A]{A. Zeisel\\
Department of Physics of Complex Systems\\
The Weizmann Institute of Science\\
Rehovot\\
Israel\\
\printead{e1}}

\address[B]{O. Zuk\\
Broad Institute of MIT and Harvard\\
Cambridge, Massachusetts\\
USA\\
\printead{e2}}

\address[C]{E. Domany\\
Department of Physics of Complex Systems\\
The Weizmann Institute of Science\\
Rehovot\\
Israel\\
\printead{e3}}
\end{aug}

\received{\smonth{9} \syear{2009}}
\revised{\smonth{8} \syear{2010}}

%
\begin{abstract}
The steep rise in availability and usage of high-throughput
technologies in biology brought with it a clear need for methods to
control the
False Discovery Rate (FDR) in multiple tests. Benjamini and Hochberg (BH)
introduced in 1995 a simple procedure and proved that it provided a
bound on the
expected value, $\mathit{FDR} \leq q$. Since then, many authors tried
to improve the BH
bound, with one approach being designing \textit{adaptive} procedures,
which aim at
estimating the number of true null hypothesis in order to get a better FDR
bound. Our two main rigorous results are the following: (i) a theorem
that provides a bound on
the FDR for adaptive procedures that use any estimator for the number
of true
hypotheses ($m_0$), (ii)~a~theorem that proves a monotonicity property of
general BH-like procedures, both for the case where the hypotheses are
independent. We also propose two improved procedures for which we prove FDR
control for the independent case, and demonstrate their advantages over several
available bounds, on simulated data and on a large number of gene expression
data sets. Both applications are simple and involve a similar amount of
computation as the original BH procedure. We compare the performance of our
proposed procedures with BH and other procedures and find that in most
cases we
get more power for the same level of statistical significance.
\end{abstract}

%
\begin{keyword}
\kwd{False Discovery Rate}
\kwd{improved BH}
\kwd{monotonicity}
\kwd{gene expression analysis}.
\end{keyword}

\end{frontmatter}

\section{\texorpdfstring{Introduction.}{Introduction}}\label{sec1}

The main goal of statistical comparisons (tests) is to calculate
the level of statistical significance at which a given null
hypothesis is rejected on the basis of available data. Researchers
use this tool in order to present their findings and support their
conclusions. Uncontrolled application of single inference
procedures in a multiple comparison setting can cause a high false
positive rate. Special multiple comparison procedures are used in
order to control the probability of committing such a type I
error in families of comparisons.

The need for improved control over the multiplicity effect in
biological experiments became acute in the nineties, when the
amount of data that could be measured and stored increased
thousands fold. Many new experimental techniques, which
allowed taking a large number of measurements simultaneously, were developed,
along with improved data acquisition and storage capabilities.

For example, in the case of gene expression microarray measurements, a
typical aim is to identify the genes whose expression levels
differentiate between healthy (type $A$) and diseased (type
$B$) subjects. Genes are tested one by one for differential expression; the
formal way to do this is by posing several thousand null
hypotheses. A null hypothesis states that a particular variable
(e.g., expression level of gene $i$) is sampled from the same
distribution for both types $A,B$; one is interested in
identifying variables (genes) for which the null hypothesis is
rejected (i.e., genes whose expression \textit{does} differentiate
between types $A,B$). Such a finding is referred to as a \textit{discovery}. Denote by $m$ the total number of hypotheses (e.g., the
number of genes whose expression levels were measured), and assume
that the null hypothesis is true for $m_0$ out of the $m$ (i.e.,
$m_0$ genes' expression levels do not differentiate the two
types). For $m_1=m-m_0$ the null hypothesis is false (the
expression levels of types $A$ and $B$ are sampled from different
distributions). A~statistical test is performed independently for
each variable, producing a~$p$-value $p_i$, $i=1,2,\ldots,m$. On the basis
of some thresholding operation on the $p_i$'s, the null hypothesis
is rejected for $R$ tests. The decision to reject (or not) can be
correct or false; When the null hypothesis is rejected for one of the $m_0$
variables for which it is actually true, we have a~``false
discovery'' (type I error). Table \ref{state_table} presents the possible
categories to which rejected and nonrejected hypotheses can
belong, and the number of hypotheses in each category.

%
\begin{table}[b]
\tablewidth=250pt
\caption{Numbers of true/false decisions taken when testing $m$
null hypotheses}\label{state_table}
\begin{tabular*}{\tablewidth}{@{\extracolsep{\fill}}lccc@{}}
\hline
\textbf{``Ground truth''} & \textbf{Nonrejected} & \textbf{Rejected} & \textbf{Total}\\
& \textbf{hypotheses} & \textbf{hypotheses} &\\
\hline
Null hypothesis is true & $U$ & $V$ & $m_0$\\
Null hypothesis is false & $T$ & $S$ & $m_1$\\
Total & $m-R$ & $R$ & $m$\\
\hline
\end{tabular*}
\end{table}

Out of the $R$ rejected hypotheses, the fraction $V/R$ is falsely
rejected. The expected value of this fraction was termed by
\citet{BH95} (referred to as BH95) as the False Discovery Rate (FDR),
%
\begin{equation} \label{eq:defFDR}
\mathit{FDR} \equiv E  \biggl( \frac{V}{R}  \Big| R >0  \biggr) \Pr(R>0)
\equiv E \biggl(\frac{V}{R^+} \biggr),
\end{equation}
where here and later in the paper the term $R^{+} \equiv
\max(R,1)$ is used for brevity. It is required since $V/R$ is
undefined when
$R=0$ and, thus,
this case should be treated separately---we follow \citet{BH95} and
replace $V/R$
by $0$ in this case.
The original BH95 procedure to control the FDR is given as
follows:
\begin{enumerate}
\label{def:BH}
\item
Denote by $q$ the desired level, $0<q \leq1$, of the FDR and
define the following set of constants:
%
\begin{equation} \label{eq:alphadef}
\alpha_i=\frac{iq}{m}, \qquad  i=1,2,\ldots,m.
\end{equation}
\item
Sort the $p$-values $p_i$ and relabel the
hypotheses accordingly, $p_{(1)}\leq p_{(2)}\leq\cdots\leq p_{(m)}$,
such that
$(i)$ is the index of the hypothesis with the $i$th smallest $p$-value.
\item
Identify $R$ as
%
\begin{equation}\label{eq:alpha}
R=\max \bigl\{i\dvtx p_{(i)}\leq\alpha_i  \bigr\}.
\end{equation}
If no such $R \geq1$ exists, no hypothesis is rejected; otherwise
reject all $R$ hypotheses $(i)=1,2,\ldots,R$.
\end{enumerate}

%
\begin{figure}

\includegraphics{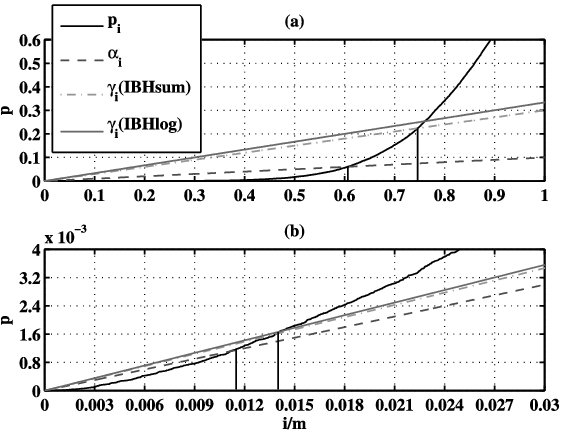}

\caption{Typical examples for the use of the BH95 and our
IBH procedures, for a desired FDR value of $q=0.1$. The sorted
$p$-values (solid line), the $\alpha_i$ of equation (\protect\ref{eq:alphadef})
(dashed line) and the $\gamma_i$ from equation (\protect\ref{def:gamma_i})
(dot--dashed line for IBHsum and solid light for IBHlog) are shown, for
\textup{(a)}~leukemia data from \protect\citet{andersson2007} and \textup{(b)} breast cancer
data from \protect\citet{pawitan2005}. As indicated in \textup{(a)}, the number of
rejections is determined for each procedure by locating the
(maximal) value $i=R$ at which the corresponding lines intersect
$p_{(i)}$ (the vertical lines mark the intersection point between the lines).}
\label{fig:example_pval}
\end{figure}

This procedure has a simple graphical implementation, depicted in Figure
\ref{fig:example_pval}. It is referred to in BH95 as ``\textit{step-up}'';
in general, there could be more than one intersection
point [of the $p_{(i)}$ and $\alpha_i$ lines], in which case the
step-up procedure identifies the intersection with the largest
$p$-value as $R$, whereas the more conservative ``\textit{step-down}''
procedure identifies the lowest one, replacing equation
(\ref{eq:alpha}) by
%
\begin{equation} \label{eq:alpha_step_down}
R=\min \bigl\{i\dvtx p_{(i)} > \alpha_i  \bigr\} - 1.
\end{equation}

The bound
%
\begin{equation}
\mathit{FDR} = E \biggl(\frac{V}{R^+} \biggr) \leq\frac{m_0}{m}q
\label{eq:BH95}
\end{equation}
was proved by BH95 for independent tests, and by \citet{BY2001} for
a certain type of ``positive dependency'' called PRDS (\textit{Positive
Regression Dependency on each one from a Subset}). The value of
$m_0$ is unknown to the researcher, but since $m_0 \leq m$, this
procedure leads
to the bound
%
\begin{equation}
\mathit{FDR} = E \biggl(\frac{V}{R^+} \biggr) \leq\frac{m_0}{m}q \leq q.
\label{eq:FDR95}
\end{equation}

Clearly, had we known $m_0$, we could have defined a different set
of constants [compare to equation (\ref{eq:alphadef})]
%
\begin{equation}
\alpha'_i=\frac{iq}{m_0} \label{eq:alphaprime}
\end{equation}
and defining
%
\begin{equation}\label{eq:alphapp}
R^\prime=\max  \bigl\{i\dvtx p_{(i)}\leq\alpha'_i  \bigr\}
\end{equation}
would have obtained a larger number $R^\prime\geq R$ of rejected
hypotheses (still with $\mathit{FDR} \leq q$) than the number $R$ given by
the original BH95 procedure, which used $m$ as an upper-bound on $m_0$.
This procedure, based on knowledge of $m_0$, is called ``oracle''
(ORC); see \citet{gavrilov2009}.
Subsequently, various improved (also called ``adaptive'') procedures
were proposed, based on the idea of estimating the unknown $m_0$ in
order to get
a more accurate handle on the FDR. These procedures can be divided into two
major classes:
\begin{enumerate}
\item Procedures for local FDR \textit{estimation}: This approach, previously
suggested and applied by
\citet{yekutieli1999}, \citet{storey2002} and \citet{poundscheng2006}, can be used
when one
has an estimator $\hatm$ of~$m_0$ that satisfies
%
\begin{equation}
m_0\leq E(\hatm)\leq m\label{eq:mhat_condition}.
\end{equation}
In procedures of this type one can write the local FDR (lFDR) estimate as
[see \citet{poundscheng2006}]
%
\begin{equation}
t_{(i)}=\frac{\hat{v}(p_{(i)})}{\hat{F}(p_{(i)})},
\end{equation}
where $p_{(i)}$ is the ordered $p$-value, $\hat{v}(\alpha)$ is the
estimator for
the type I errors (in the rejection region), and $\hat{F}(\alpha)$ is the
estimator for the probability $\Pr(p\leq\alpha)$ [often estimated for $p_{(i)}$
by $i/m$]. Since for $\hat{v}(\alpha)$ most methods use
%
\begin{equation}
\hat{v}(\alpha)=\alpha\frac{\hatm}{m},
\end{equation}
any estimator that satisfies equation (\ref{eq:mhat_condition}) can
provide an
improved estimator for the local-FDR by
%
\begin{equation}
\alpha\frac{m_0}{m}\leq E(\hat{v}(\alpha))=\alpha\frac{E(\hatm
)}{m}\leq\alpha.
\end{equation}
When using $\hat{F}(p_{(i)}) = i/m$ one gets an improved bound on the local-FDR
estimator:
%
\begin{equation}
\frac{p_{(i)}}{i} m_0 \leq E \bigl(t_{(i)}\bigr) = \frac{p_{(i)}}{i} E(\hatm)
\leq
\frac{p_{(i)}}{i} m.
\end{equation}
This approach is the preferred one in many biological contexts when the
investigator wishes to control $R$, the number of discoveries made (e.g.,
differentiating genes to be used in further experiments).
%
\item Procedures for FDR \textit{control}: In this approach, one
wishes to control the FDR at a preset level $q$. This is achieved by
defining $\gamma_i=iq/\hatm$ to be used in the same way as
$\alpha_i$ and $\alpha_i^\prime$ [see equations (\ref{eq:alphadef})
and (\ref{eq:alphaprime})], leading typically to a larger number $R^\prime
$ of
rejected hypotheses (compared to BH95), with the $\mathit{FDR}$ still being
bound by the desired value $q$. The advantage of this procedure
(presented in Section \ref{sec:synthetic_data}) is that one retains
control of $q$, the desired level of FDR.

\end{enumerate}

We present in this paper two estimators, $\hatm$ and $\galm$, that satisfy
equation (\ref{eq:mhat_condition}), and hence can be used trivially
for FDR
estimation. As opposed to FDR estimation, proving \textit{control} of
the FDR is
far more involved, and constitutes a significant portion of this paper. We
provide two new proven procedures for control of the FDR. We first
prove control
for these procedures when employed in a step-up manner. Then, by using
a new
general monotonicity result for the FDR which we derive, we show that the
step-down versions of our procedures also control the FDR. Designing better
procedures for FDR estimation and control has drawn a great deal of
attention in recent years, as is demonstrated by the abundance of proposed
procedures and many theoretical and experimental papers. However, as
far as we
know, only for a few such procedures has control of the FDR been rigorously
established: the original BH95 procedure \citet{BH95}, the two-stage and
multiple-stage adaptive BH procedures \citet{BKY} (we refer to the
latter as
BKY), and Storey's procedure \citet{storey2002} (referred to as STS).
All these
procedures (except, of course, BH95) claim to give improved power over
BH95. All
are derived from a better estimation of $m_0$. Almost all proofs for
FDR control
assume independence of the $p$-values [with the notable exception of
\citet{BY2001}].
Thus, far less is known about the behavior of FDR procedures under dependency,
where most of our understanding comes from simulation studies. In addition,
the FDR, by its definition [equation (\ref{eq:defFDR})], is an \textit{expected}
value. However, the fraction of the false discoveries $V/R^+$ is a
\textit{random variable}. While the mean value (FDR) was extensively
studied, far
less attention has been devoted in the literature to the behavior of
this random
variable, its variance and entire distribution. We therefore perform simulations
whose purposes are as follows: (a) To study the behavior of the various
procedures under
dependence, where analytical results are harder to establish, and
(b) study the distribution of the fraction of false rejections ($V/R^+$),
which has implications on possible violation of the bound for a
particular realization. Our simulations provide a comparison of our new
procedures to the known ones mentioned above and we show that our new procedures
compare favorably in most cases of interest. We analyze simulated and real
data, and show that for both the new procedures almost always
reject more hypotheses than BH95, while maintaining control even under
dependence, and we therefore refer to these procedures as ``Improved BH'' (IBH).
The real data which we use is gene expression data obtained from
various cancer studies, and we show that our new procedures allow
rejection of
more hypotheses at a given confidence level and thus increase discovery power.

A Matlab package implementing our proposed procedures, including\break
\mbox{examples} and
data sets analyzed in the paper, is provided in the supple-\break mentary information
\citet{ZZDsupp2010} and
in the fol-\break lowing URL:
\href{http://www.broadinstitute.org/\textasciitilde orzuk/matlab/libs/stats/fdr/matlab_fdr_utils.html}%
{http://www.broadinstitute.org/\textasciitilde
orzuk/matlab/libs/stats/fdr/}\break
\href{http://www.broadinstitute.org/\textasciitilde orzuk/matlab/libs/stats/fdr/matlab_fdr_utils.html}%
{matlab\_fdr\_utils.html}.

\section{\texorpdfstring{Preliminaries and theorem on control.}{Preliminaries and theorem on control}}
In this section we present a~theorem which provides a general way
to build an improved bound for controlling the FDR using an
estimator for $m_0$. Two examples of practical implementation of
the theorem lead to useful procedures described in the next
section. The working assumption we use here is that the $p$-values are
independent. The theorem is not proven for dependent variables, but our
simulations indicate that in most cases we do control the FDR even under
dependence (see Section \ref{sec:synthetic_data}).
Our first step is defining mathematically a family of estimators $\hatm$
for $m_0$. We define a general modified BH procedure, in which any one
of these estimators is used by replacing $m$ in the original BH95
procedure [see
equations (\ref{eq:alphadef}) and (\ref{eq:alpha})] by $\hatm$. Throughout
this section and
the rest of the paper we denote for convenience $p_{i..j} \equiv
p_i,\ldots,p_j$. We also denote $\vecp= (p_1,\ldots,p_m)$ the vector of
all $p$-values.
\begin{definition}\label{def:monotonit}
An estimator for $m_0$ is a family of functions $\hatm\equiv
\hatm^{(m)}\dvtx [0,1]^m \to\mathbb{R}$, $\hatm\equiv\hatm(\vecp)$.
We usually omit the index $^{(m)}$, as it is obvious from the
context. We say that $\hatm$ is a monotonic estimator if it
satisfies the following:
\begin{enumerate}
\item$\hatm^{(m)}(p_1,\ldots,p_i,\ldots,p_m) \!\geq\!
\hatm^{(m)}(p_1,\ldots,p_i',\ldots,p_m)$,  $\forall   p_i \geq
p_i'$,  $i=1,2,\ldots, m$,   $  m \geq1$.
\item
$\hatm^{(m)}(p_1,\ldots,p_i,\ldots,p_m) \geq
\hatm^{(m-1)}(p_1,\ldots,p_{i-1},p_{i+1},\ldots,p_{m})$, $ \forall
i=1,2, \ldots,\break m$,  $ m \geq2$.
\end{enumerate}
\end{definition}

\begin{definition}\label{def:modified_BH}
Assume w.l.o.g that we have $m$ hypotheses, the first~$m_0$ of which
are null. Let $\vecp= (p_1,\ldots,p_m)$ be the corresponding $p$-values.
The modified step-up BH procedure with estimator $\hatm$ is defined as follows:
\begin{enumerate}
\item Compute $\hatm\equiv\hatm(\vecp)$.\vadjust{\goodbreak}

\item For each $i$ define
%
\begin{equation}
\gamma_i = \frac{i q}{\hatm} \label{def:gamma_i}.
\end{equation}
\item Order the $p$-values in an increasing order: $p_{(1)} \leq\cdots
\leq p_{(m)}$.

\item
Let $R = \max\{i\dvtx p_{(i)} \leq\gamma_i\}$, and
reject the hypotheses $(1),(2),\ldots,(R)$ (if no such $R$ exists, do not
reject any hypothesis).
\end{enumerate}
This procedure is similar to the original BH95 procedure, with the
additional initial step of estimating $m_0$, and the different set of
constants used to determine $R$. The modified step-down BH
procedure is defined in the same way, except that in step $4$ we take
$R=\min \{i\dvtx p_{(i)} >\alpha_i  \} - 1$.
\end{definition}

The next theorem gives the bound on the FDR for the above
procedure under the above assumptions [a very similar result was
given by \citet{BKY}].

\begin{thm}\label{thm:m0_control}
$\!\!\!$Let $\hatm\equiv\hatm(\vecp)$ be a monotonic estimator for
$m_0$. Consi\-der the modified step-up BH procedure defined above. Let
$\hatmone(\vecp) \equiv\hatm(p_2,\ldots,\break p_m)$ be the same estimator,
but disregarding the first (null) $p$-value $p_1$. Assume that the null
$p$-values are i.i.d. $U[0,1]$. Then the procedure satisfies
%
\begin{equation}
\mathit{FDR} = E \biggl[\frac{V}{R^{+}} \biggr] \leq m_0 q
E \biggl[\frac{1}{\hatmone} \biggr].
\end{equation}
\end{thm}

Here $p_1$ is a representative of one of the true null $p$-values.
The modified estimator $\hatmone$ which excludes $p_1$ cannot be implemented
in practice, as the researcher does not know which of the $p$-values are null,
but for any estimator $\hatm$ we can still consider this hypothetical estimator
(in similar
vain to the ``oracle'' procedure sometimes considered in the literature)
and study
its statistical properties---it only serves for a hypothetical auxiliary
procedure which is used in the proof
of the theorem, and the theorem applies to the practical original
procedure with
the
estimator $\hatm$ which does use $p_1$ (as well as all other $p$-values).
The proof of Theorem \ref{thm:m0_control} is given in
\citet{ZZDsupp2010}, Supplement A for
completeness. In general, a direct computation, or bounding of the FDR
for a
given procedure, is a demanding task, which depends heavily on the
procedure's details, and suffers from complicated dependence on the
rejection of different hypotheses, reflected in the computation of
$E[V/R^+]$ (this is true even if the $p$-values themselves are
independent) and, therefore, there is no general way to prove FDR
controlling properties of various procedures. The advantage of Theorem~%
\ref{thm:m0_control} is that it provides a direct method for proving
control for
a wide
class of procedures, by simply bounding the reciprocal mean of the
estimator for $m_0$. In the next section we use this theorem to prove control
of the FDR for two procedures, based on different estimators $\hatm$
and $\galm$ which we propose. We are not aware of a direct way for proving
control of the FDR for these procedures, thus demonstrating the power and
generality of the theorem.

\section{\texorpdfstring{The proposed procedures.}{The proposed procedures}}\label{sec:proposed_procedure}
In this section we propose two FDR controlling procedures. We show
that they achieve direct control of $q$, the desired value of the
FDR, while producing a list of $R'$ discoveries satisfying almost
always $R' \geq R$, the corresponding BH95 value. The procedures
are particular cases of Definition~\ref{def:modified_BH}. According to Theorem
\ref{thm:m0_control}, any estimator that satisfies our monotonicity assumption
bounds the FDR by $\mathit{FDR}\leq m_0 q E [1/\hatmone]$.
Therefore, in order to show that the FDR is controlled, it suffices to
bound $E [1/\hatmone]$. In particular, if we want to achieve a certain FDR
control level~$q$, we need to verify that
%
\begin{equation}
E \biggl[\frac{1}{\hatmone} \biggr]\leq\frac{1}{m_0}.
\label{eq:inv_mean_m0_cond}
\end{equation}

Our first estimator is based on
%
\begin{equation}
\hatm^\prime= 2\sum_{j=1}^m p_j \label{eq:mhatp}.
\end{equation}
$\hatm^\prime$ was used by \citet{poundscheng2006} for estimation, but
without proving control of the FDR. The second estimator
is based on
%
\begin{equation}
\galm^\prime=-\sum_{i=1}^m \log(1-p_i) \label{eq:galmp}.
\end{equation}

For both estimators we first show that equation (\ref
{eq:mhat_condition}) is
satisfied and, hence, both can be used for FDR \textit{estimation}. Next we
describe the procedure to be used for \textit{control} of the FDR, which is
proved by showing, for slightly modified versions of both
estimators (see below), $\hatm$ and $\galm$, that the bound equation
(\ref{eq:inv_mean_m0_cond})
is satisfied. Both $\hatm^\prime,\galm^\prime$ are monotonic
estimators according to Definition \ref{def:monotonit}. Our claims are
as follows:
\begin{enumerate}
\item Both estimators are conservative, that is, their expectation is
at least~$m_0$. Moreover, as the
statistical power of each individual test increases, and the $p_i$ of the
alternative hypothesis approach zero, our estimators converge (in expectation)
to the true
value of $m_0$.
\item Both procedures control the FDR---for the list of $R'$
discoveries we have \mbox{$\mathit{FDR} \leq q$}.
\item In nearly all cases of interest the number of discoveries
obtained by our procedures exceeds the number obtained (for the
same value of $q$) by the BH95 procedure, that is, $R' \geq R$.
This holds since nearly always $\hatm\leq m$ (exceptions occur
when there are almost no false hypotheses, that is, $m$ and $m_0$ are
very close).

\end{enumerate}

A reasonable requirement from an estimator for $m_0$ should be that it is
conservative (i.e., larger than $m_0$ in expectation). We would also like
our estimator to be (approximately) unbiased, at least when all
hypotheses are
null, since otherwise we will get a systematic overestimation of $m_0$
and a~corresponding underestimation of the FDR. Finally, a desirable property
is being \textit{asymptotically unbiased}---that is, even when there are nonnull
hypotheses, when the sample size of the individual tests grows to infinity,
we would want the estimator to converge, on expectation, to the true
value~$m_0$.
These properties were dealt with in \citet{poundscheng2006}, where it
was shown
that~$\hatm^\prime$ indeed satisfy them. Here we show them for both our procedures:

\begin{claim}
\textup{(a)} Both estimators are conservative:
%
\begin{equation}
E [\hatm^\prime]\geq m,\qquad  E[\galm^\prime] \geq m_0.
\end{equation}

\textup{(b)} Assume that the sample size of all tests goes to infinity, and,
thus, $E
[p_i] \to0$ for $i=m_0+1,\ldots,m$.
Then both estimators converge in expectation to $m_0$:
%
\begin{equation}
E[\hatm^\prime] \to m_0, \qquad   E[\galm^\prime] \to m_0.
\end{equation}
\end{claim}

\begin{pf}
(a)
%
\begin{eqnarray}
E [\hatm^\prime] &=& 2 \sum_{j=1}^m E[p_j] = 2 \Biggl(\sum_{j=1}^{m_0}
E[p_j] +
\sum_{j=m_0+1}^m E[p_j]\Biggr)\nonumber\\[-8pt]\\[-8pt]
& =&
m_0 + 2 \sum_{j=m_0+1}^m E[p_j] \geq m_0 ,\nonumber
\\
\hspace*{10pt}E [\galm^\prime] &=& \sum_{j=1}^m E[\log(1-p_j)] =
\sum_{j=1}^{m_0} E[\log(1-p_j)] + \sum_{j=m_0+1}^m
E[\log(1-p_j)]\hspace*{-8pt}\nonumber\\[-8pt]\\[-8pt]
& =&m_0 + \sum_{j=m_0+1}^m E[\log(1-p_j)] \geq m_0.\nonumber
\end{eqnarray}

(b) From the two equations above it is clear that as all the alternative
$E[p_j]$ approach zero, the expectation of both estimators converges to $m_0$.
\end{pf}

In order to show control of the FDR using Theorem \ref
{thm:m0_control}, we have to
apply
small corrections to both estimators, turning them
into conservative estimators (i.e., overestimating $m_0$). This is
due to two reasons: the first is that the bound on the FDR given in Theorem
\ref{thm:m0_control} uses $\hatmone$ (rather than $\hatm$) and, thus, we
``lose'' one of
the $p$-values and need to correct for that. The second reason is that
$\hatmone$
appears in the denominator, and its fluctuations have asymmetric
influence on
the FDR bound. This can be illustrated by using Jensen's inequality
which gives $E[1/ \hatmone] \geq1 / E[\hatmone]$, thus showing that
an unbiased estimator for $m_0$ will typically show a bias when its
reciprocal is used. Nevertheless, we show that these two effects can
be overcome by applying a small correction, which becomes negligible
as the number of hypotheses goes to infinity.

\subsection{\texorpdfstring{The IBHsum procedure.}{The IBHsum procedure}}
Our first estimator is based on $\hatm^\prime$ [see equation~(\ref{eq:mhatp})] that was also used by \citet{poundscheng2006}, but
only for \textit{estimation} and not for \textit{control}. Since for the
$m_0$ variables for which the null hypothesis holds we have
$p_i^{\mathrm{true}}\sim
U[0,1]\Rightarrow E[p_i^{\mathrm{true}}]=\frac{1}{2}$, it is trivial to see that
$E[\hatm^\prime]\geq m_0$. To show that $E[\hatm^\prime]\leq m$,
we have to make a further assumption regarding the alternative
$p$-values $p_i^{\mathrm{false}}$: We denote the distribution of $p_i^{\mathrm{false}}$ by
$f_i^{\mathrm{false}}$, that is, $p_i^{\mathrm{false}} \sim f_i^{\mathrm{false}}$. If all
the $f_i$'s are
\textit{stochastically smaller} [\citet{Aven01}] than the uniform distribution
($f_i^{\mathrm{false}} \leq_{\mathrm{st}} U[0,1]$), we have $E[p_i^{\mathrm{false}}]\leq\frac
{1}{2}$ which
immediately implies $E[\hatm^\prime]\leq m$ [a probability density function
$f$ is
said to be stochastically smaller than a probability density function $g$,
$f \leq_{\mathrm{st}} g$, if $F(x)=\int_{-\infty}^{x}f(t)\,dt \geq
G(x)=\int_{-\infty}^{x}g(t)\,dt $ $\forall x \in(-\infty,\infty);$
\citet{Aven01}].
%
\begin{figure}[b]

\includegraphics{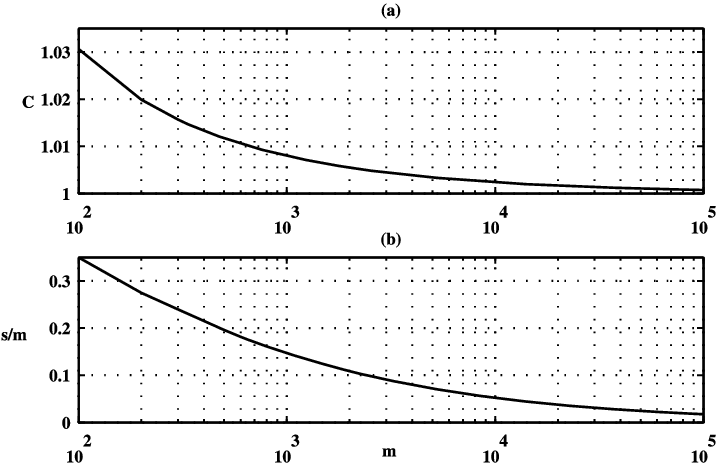}

\caption{The correction functions $C(m)$ and $s(m)/m$ [see equation
(\protect\ref{eq:Chatm0})]. As $m \to\infty$ the multiplicative correction
$C(m)$ approaches one, while the (normalized) threshold $s(m)/m$ [used when
$\hatm' \leq s(m)$] goes to zero, thus, $\hatm$ reduces to the uncorrected
$\hatm'$.}\label{fig:C_s_vs_m}
\end{figure}

We introduce the following modified estimator:
%
\begin{equation}
\hatm=C(m)\cdot \min [m, \max (s(m),\hatm^\prime ) ],
\label{eq:Chatm0}
\end{equation}
where $C(m),s(m)$ are universal correction factors that ensure that
the condition~(\ref{eq:inv_mean_m0_cond}) is satisfied [for details
see \citet{ZZDsupp2010}, Supplement~B]. The correction factors were computed
numerically and are presented in Figure~\ref{fig:C_s_vs_m}. When
$m\rightarrow\infty$, $C \rightarrow1$ and $s/m \rightarrow0$,
and, therefore, the corrections become negligible and the estimator
$\hatm$
reduces to $\hatm^\prime$.

\subsection{\texorpdfstring{The IBHlog estimator.}{The IBHlog estimator}}
In this section we propose another estimator for $m_0$, based on $
\galm^\prime$ [see equation (\ref{eq:galmp})]. Again, since for
$i=1,2,\ldots,m_0$ we have $p_i^{\mathrm{true}} \sim U[0,1] \Rightarrow E[-\log(1-p_i)]=1$
and, therefore, $E[\galm^\prime]\geq m_0$. Furthermore, if all the
alternative $p$-values $p_i^{\mathrm{false}}$ have a distribution which is
stochastically smaller than the uniform distribution
$(f_{p_i}^{\mathrm{false}}(p) \leq_{\mathrm{st}} U[0,1])$, then
$E[-\log(1-p_i^{\mathrm{false}})]\leq1$, and, therefore,
$E[\galm^\prime]\leq m$.

The advantage of using the second estimator $\galm^\prime$ over
$\hatm^\prime$ is that when $f_{p_i}^{\mathrm{false}}(p) \leq_{\mathrm{st}}
U[0,1]$, the alternative hypothesis generates $p$-values skewed to
the left. Since $-\log(1-p) < 2p$,  $ \forall p < \frac{1}{2}$
[see equations (\ref{eq:mhatp}) and (\ref{eq:galmp})], this typically implies
$\galm^\prime\leq\hatm^\prime$ and, thus, $\galm^\prime$ is
typically closer to
the true $m_0$.
A possible drawback is that the variance of $\galm^\prime$ is
typically larger than that of $\hatm^\prime$, which might result in
an instability in the estimation of $m_0$.

Proving control of the FDR for $\hatm$ is difficult since we need to
bound $1 /
\hatm$ which has a complicated distribution. Here we show that the distribution
of $\galm^\prime$ is much simpler, and this enables us to prove
control of the FDR by introducing only a slight additive correction.

\begin{claim}
Define the (corrected) estimator:\vspace*{-2pt}
%
\begin{equation}
\galm\equiv2+\galm^\prime= 2-\sum_{i=1}^m \log(1-p_i).
\label{def:mgal}\vspace*{-2pt}
\end{equation}
Assume that the null $p$-values are i.i.d.
$U[0,1]$. Then the modified $BH$ procedure with estimator $\galm$ and
parameter $q$ controls the FDR at level $\leq q$.
\label{logp_estimator_claim}\vspace*{-2pt}
\end{claim}

The proof is achieved by bounding $E[1/\galmone]$ and then using
Theorem~\ref{thm:m0_control}. See \citet{ZZDsupp2010}, Supplement C
for full details.

\section{Is the FDR monotonic?}\label{sec:monotoncity_FDR}
In this section we take a slight detour from the study of our
proposed procedures to investigate the following question: is it
generally true that by modifying an FDR procedure to be more
stringent, one is guaranteed to obtain a more conservative control on the
FDR? The motivation for dealing with this question in the context of
the current
paper (which deals with the control property of a modified BH
procedure) comes
from the fact that Theorem \ref{thm:m0_control} was proved only for step-up
procedures, which leads us to ask whether it holds also for the more
conservative step-down case. Monotonicity is a natural property that
one might
expect when performing statistical tests, as it allows the researcher
to choose
a trade-off between maximizing the statistical power and minimizing the
risk of
making false discoveries. The analogous question for a single
hypothesis is
whether taking a more conservative (lower) $p$-value cutoff guarantees to reduce
the risk of making a type-I error, and is trivially answered in the
affirmative. Our formulation of the question in the multiple-hypothesis
settings using FDR is as follows: Given two procedures, $B^{(1)}, B^{(2)}$
(possibly parameterized by $q$ or other parameters), and assuming
that for any realization of the $p$-values, $B^{(2)}$
passes more hypotheses than $B^{(1)}$, is it true that $\mathit{FDR}^{(1)} \leq
\mathit{FDR}^{(2)}$? While this statement seems a natural and plausible
property of FDR procedures, we are not aware of any previous
treatment of it in the literature. Here we show that under certain
monotonicity conditions on the alternative hypothesis $p$-values distribution,
one can prove this monotonicity property of the FDR.

%
\begin{thm}\label{thm:monotone}
Let $\vecp= (p_{1,\ldots,m})$ be a set of independent $p$-values.
Assume that $f$, the marginal probability density function of the alternatives,
is monotonically nonincreasing and differentiable. Let $B^{(i)}$ be two
threshold FDR procedures rejecting $R^{(i)}(\vecp)$ hypotheses and
each having
$\mathit{FDR}^{(i)}$, $i=1,2$.
Assume that for any $q$, $R^{(1)}(\vecp) \leq R^{(2)}(\vecp),
\forall\vecp$.
Then it also holds that $\mathit{FDR}^{(1)} \leq \mathit{FDR}^{(2)}$.
\end{thm}

The proof is given in \citet{ZZDsupp2010}, Supplement D.
A particular application of the above theorem is showing that
step-down procedures give better FDR then step-up procedures.
Thus, we immediately get the following:
\begin{corollary}
The statement of Theorem \ref{thm:m0_control} holds also for the
step-down procedure, provided that the alternative $f$ is
monotonically decreasing.
\end{corollary}

The above conditions for monotonicity might appear a bit restrictive,
and one could hope to relax them---for example, require only
$f \leq_{\mathrm{st}} U[0,1]$ instead of monotonicity. We have found that,
perhaps surprisingly, monotonicity of the FDR does not hold under such
relaxed conditions, by giving an example in which FDR monotonicity is violated,
even for a simple case of independent test statistics (both null and
nonnull), when $f \leq_{\mathrm{st}} U[0,1]$, and when the FDR procedures
themselves are
monotonic. It is thus not obvious at all that in practice we will
always observe
a monotonic behavior of the FDR, and, thus, it is possible to get a
higher FDR for a more conservative procedure.

\begin{example}
Let $m=3$ and $m_0=1$. Let the two alternative hypotheses $p$-values
be taken from a mixture distribution, $p_i \sim\epsilon
U[0,\epsilon]+(1-\epsilon)\delta(p_i-\epsilon)$
for some $0 < \eps< 1$. Thus, $p_2, p_3$ are ``truncated'' uniform
r.v.s., having
$1-\eps$
of their mass concentrated at $\eps$, and the rest ($\eps$) uniformly
distributed on $[0,\eps]$; their distributions are stochastically
smaller than
$U[0,1]$. For simplicity of computations, we assume that $\eps\ll 1$
and thus
look only at the first order in $\eps$, although the example's
conclusion holds
for any $\eps> 0$. Let $P^{(1)}$ be the procedure always rejecting the lowest
$p$-value and $P^{(2)}$ be the procedure rejecting the two lowest
$p$-values (we
assume that ties are handled in the same way by both procedures, for
example, by taking
$p$-values in lexicographic order---the precise tie-breaking rule does
not change
the example's results). We next compute the FDR for both procedures:
%
\begin{eqnarray}
\mathit{FDR}^{(1)} &=& \Pr(p_1 < p_2, p_3) = \eps[ \eps^2 / 3 + 2
\eps(1-\eps) / 2 + (1-\eps)^2 ] \nonumber\\[-8pt]\\[-8pt]
&=& \eps+ O(\eps^2),\nonumber
\\
\mathit{FDR}^{(2)} &=& \bigl(1 - \Pr( p_1 > p_2, p_3)\bigr) / 2 = [1 -
(1-\eps) -
\eps^3/3]/2 \nonumber\\[-8pt]\\[-8pt]
&=& \eps/2 + O(\eps^3).\nonumber
\end{eqnarray}
Thus, for $\eps$ small enough $\mathit{FDR}^{(1)} > \mathit{FDR}^{(2)}$ and the more
conservative procedure leads, in fact, to a higher FDR.
\end{example}

\section{\texorpdfstring{Synthetic data obtained by simulations.}{Synthetic data obtained by simulations}}\label{sec:synthetic_data}
We applied our method, as well as several others (see below), to
synthetic data
obtained by simulations performed along the lines of \citet{gavrilov2009}, with
full details presented in \citet{ZZDsupp2010}, Supplement E. The
advantage of
working with synthetic data is that several parameters of interest are under
full control, and one can investigate their effect on the quality of different
procedures and bounds. Furthermore, by performing repeated simulations,
one can
determine not only the (\textit{expected} value) FDR but also the entire
\textit{distribution} of $V/R^+$. One should bear in mind that results based on
specific simulations might have limited applicability and are hard to
generalize, since the simulations use specific configurations (e.g., data
distribution, test to determine $p$-values, hypothesis dependency structure,
etc.). A comprehensive simulation capturing all possible behaviors of the
hypothesis is infeasible, but we have tried to explore various different
plausible scenarios which might be encountered in practice, by changing the
number of (total and null) hypotheses and their dependency structure,
with both positive and negative correlations. The simulations produce
two kinds
of Gaussian random variables: $Z_1,\ldots,Z_{m_0}$, sampled from the
standard normal
distribution $P_0 \equiv N(0,1)$, and $Z_{m_0+1},\ldots,Z_m$, sampled from $P_1
\equiv N(\mu_1,1)$, centered on $\mu_1>0$. All variables (both null and
nonnull) are sampled with covariance~$\rho$ ($0 \leq\rho\leq1)$:
at the
extreme cases, setting $\rho= 0$ corresponds to independent variables, whereas
$\rho= 1$ to full (deterministic) dependency. For each~$Z_i$ the corresponding
\textit{two-tailed} $p$-value is obtained, $p_i=2 \Phi(-\vert Z_i \vert)$,
where~$\Phi$ is the standard Gaussian cumulative distribution function. The obtained
$p_i$'s have a uniform $U[0,1]$ distribution for $i=1,\ldots,m_0$
(corresponding to
the null hypothesis) and a distribution stochastically smaller than
uniform for
$i=m_0+1,\ldots,m$ (the alternative hypothesis).

A set of $m$ such variables constitutes a single instance or
realization of the data to be analyzed. To get accurate estimates
of the FDR and the $V/R^+$ distribution, we generated for each simulation
50,000 such realizations, which generally gave highly accurate and reproducible
estimates. Under the null hypotheses all variables are sampled from the first
distribution, $m$ $p$-values are calculated accordingly and used as input
to one
of the procedures with a desired FDR bound $q$, producing a list of
$R$ rejections. As opposed to real data, here one can go back and
identify those $V$ among the~$R$ that were falsely rejected (i.e.,
were, in fact, selected from $P_0$). This way one can keep track
of the true values of $V/R^+$, their mean (calculated over a~large
number of instances), variance, etc.
One important goal of the simulation is comparing our procedures
to existing ones. Specifically, we compare our procedure to the
following: (1) the BH95
procedure as described in the \hyperref[sec1]{Introduction}, (2) the BKY procedure which defines
a local ($i$-dependent)\vadjust{\eject} estimator for~$m_0$, given by
$\hat{m}_0^{\mathrm{BKY}}=m+1-i(1-q)$, and uses it in the step-down manner of
the BH95 procedure with $q^*=qm/\hat{m}_0^{\mathrm{BKY}}$, (3) the STS
procedure which introduces $\hat{m}_0^{\mathrm{STS}}=(m+1-r(\lambda
))/(1-\lambda)$ as the
estimator for~$m_0$ where $r(\lambda)=\#\{p_i\leq\lambda\}$, and
then uses the
step-up BH95 procedure, with $q^*=qm/\hat{m}_0^{\mathrm{STS}}$, with the
requirement that all the rejected $p_i\leq\lambda$ (throughout this
paper we used the STS procedure with $\lambda=0.5$).
We present here two kinds of results derived from such
simulations. First we compare the values of $\mathit{FDR} = E(V/R^+)$ obtained
by the
procedures discussed above: BH95, BKY, STS, IBHsum and IBHlog when the
hypotheses are dependent. In particular, we
demonstrate that for positive correlations $\rho>0$ our IBH as well as
the BKY
procedures yield, for a given desired value of $q$, an
FDR that is either less than $q$ or exceeds it slightly.
On the other hand, the STS method produces, for $\rho>0$, values of
FDR that
exceed $q$ by a large margin. The second aim is
to assess the extent to which the value of $V/R^+$, obtained for a~%
particular realization, will violate the bound, especially for the
IBH methods.\looseness=-1

\begin{figure}

\includegraphics{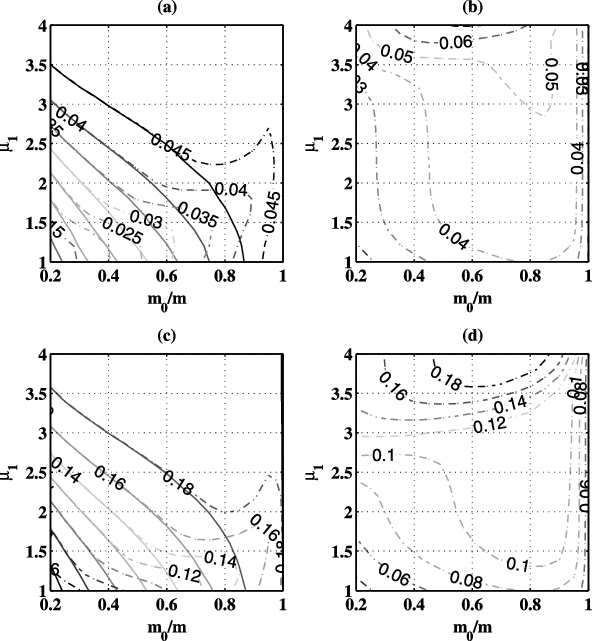}%
\vspace*{-5pt}
\caption{Isolines of $E(V/R^+)$, measured for the IBHsum procedure
by simulations, presented in the $(\mu_1,m_0/m)$ plane. The solid
lines in \textup{(a)}
and \textup{(c)} are for the step-up procedure and the dashed lines for the step-down
procedure. \textup{(a)} and \textup{(c)} are for the independent case ($\rho=0$).
\textup{(b)} and \textup{(d)} are for the positive dependency case ($\rho=0.8$).
The FDR levels are $q=0.05$ in \textup{(a)}, \textup{(b)} and $q=0.2$ in \textup{(c)}, \textup{(d)}. In \textup{(b)}
we find
$E(V/R^+)> 0.05$ for large $\mu_1$, in violation of the bound
$q=0.05$. The
step-up and step-down procedures tend to coincide for independent
$p$-values and
low $m_0/m$; the differences between them are more significant when the signal
is weak (small $\mu_1$) and $m_0/m$ is high.
\label{fig:m_500_q5_q20_rho_0and0p8_EVR_maps}}
\vspace*{-7pt}
\end{figure}

As an overview we start by presenting in Figure
\ref{fig:m_500_q5_q20_rho_0and0p8_EVR_maps} the performance of our
proposed IBHsum procedure for fixed $m=500$ and $q=0.05,0.2$, and for
a wide range of the parameters $m_0 / m$ (fraction
of alternative hypotheses) and $\mu_1$ (signal strength), by
estimating the
expected value $\mathit{FDR} = E(V/R^+)$ from our simulations.
Figure \ref{fig:m_500_q5_q20_rho_0and0p8_EVR_maps}a and c are for the
independent case and show both step-down and step-up results. As we
can see, the two become identical when the signal ($\mu_1$) is
strong or when $m_0/m$ is small. Figure
\ref{fig:m_500_q5_q20_rho_0and0p8_EVR_maps}b and d are for the
positively dependent case ($\rho=0.8$) for which the procedure is
not proved to control the FDR. Indeed, we can observe in Figure
\ref{fig:m_500_q5_q20_rho_0and0p8_EVR_maps}b violation of the FDR
level $q$ for large signals ($\mu_1$); this violation of the bound for the
dependent case will be discussed later.

\subsection{\texorpdfstring{Comparison of several methods under dependency.}{Comparison of several methods under dependency}}\label{subsec:comparison_of_methods}
\begin{figure}

\includegraphics{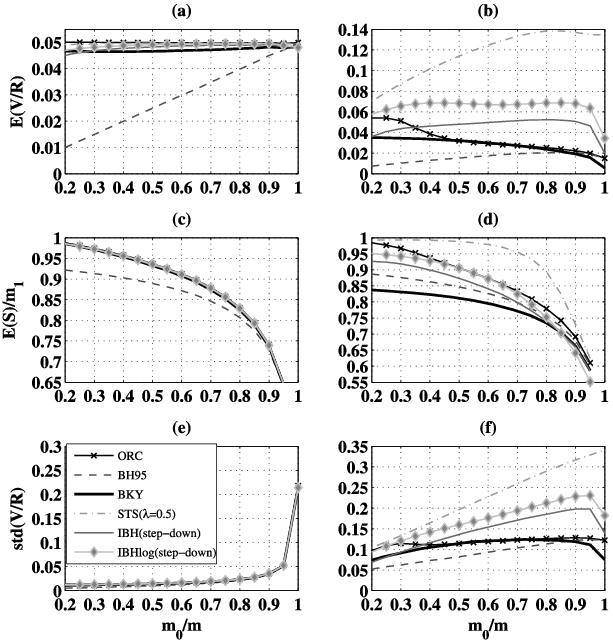}

\caption{Results obtained for synthetic data with $m=500$
hypotheses; $m_0$ was varied, the FDR was set at $q=0.05$, the mean
of the distributions $P_1$ was $\mu_1=3.5$ and the data were drawn
either with covariance $\rho= 0$ [\textup{(a)}, \textup{(c)} and \textup{(e)}] or $\rho= 0.8$ [\textup{(b)},
\textup{(d)} and \textup{(f)}]. Six methods were compared: oracle (ORC), BH95, BKY, STS
and our two IBH procedures (in a step-down manner), showing
$E(V/R^+)$ in \textup{(a)} and \textup{(b)}, the power $E(S)/m_1$ in \textup{(c)} and \textup{(d)}, and
the standard deviation (st.d.) of $V/R^+$ in \textup{(e)} and \textup{(f)}, for the
independent case and positively dependent cases, respectively.
\label{fig:comparison_EVR_and_ESm}}
\end{figure}

Here we fixed the signal parameter $\mu_1 = 3.5$, and varied $m_0/m$
between 0.2
and 1 (for $m=500$). We present, in Figure \ref
{fig:comparison_EVR_and_ESm}a, c
and e, results obtained for $\rho= 0$ (complete independence) and in Figure
\ref{fig:comparison_EVR_and_ESm}b, d and f for $\rho= 0.8$ (strong
dependence).
For each instance we applied the five procedures with $q=0.05$. For STS we
chose $\lambda= 0.5$, and our IBHsum and IBHlog were employed in a step-down
manner. Figure \ref{fig:comparison_EVR_and_ESm}a and b present for
each method
the mean value of $V/R^+$, as a function of $m_0/m$. These means provide
excellent estimates of $E(V/R^+)$, and they reveal that, as expected,
for $\rho
= 0$ all methods satisfy the bound $E(V/R^+) \leq q$. The STS and IBH come
closest to saturating the bound, with BKY slightly lower and BH95 significantly
lower. The figures show also the result obtained by an ``oracle,''
namely, the
procedure that uses the known value of $m_0$ in order to determine
$R'$ according to equations~(\ref{eq:alphaprime}) and (\ref{eq:alphapp}).

For $\rho> 0$ no proved upper bound exists for either of the BKY,
STS or IBH procedures. Furthermore, the proof of \citet{BY2001} for
the BH95 procedure does not hold for two-tailed tests: indeed, as
can be seen on Figure \ref{fig:comparison_EVR_and_ESm}b, the FDR
obtained by the oracle procedure (slightly) violates the bound
$q=0.05$ for $m_0/m \leq0.3$, in agreement with the violation reported
in \citet{Reiner2007}. Therefore, it is important to assess
the extent to which $E(V/R^+)$ obtained by each of these methods
violates the bound~$q$ in the presence of positive correlations
between the hypotheses. As seen in Figure
\ref{fig:comparison_EVR_and_ESm}b, for
$\rho=0.8$ the STS method produces a measured FDR that
overshoots the value $q=0.05$ of the bound by more than twice, for
most of the range of $m_0$ values studied. In comparison, the other
methods (BH95, BKY, IBHsum) provide FDR which remains below the bound
or exceeds it slightly for a narrow range of $m_0$. The IBHlog
procedure also violates the bound for nearly the entire range of
$m_0/m$, but by much less than STS.

We conclude these comparisons between the different procedures by
presenting, in Figure \ref{fig:comparison_EVR_and_ESm}c and d,
their power, measured as the fraction of correctly rejected
hypotheses, or ``True Discovery Rate.'' For each realization we
calculated $S=R-V$ and plotted the ratio $S/m_1 = (R-V)/(m-m_0)$,
averaged over all instances. This measure of power is one minus
the type two error rate, known as the False Non-Discovery Rate
$T/m_1$ [\citet{Genovese2002}]. For the independent case $\rho= 0$
the power values of the ORC, BKY, STS and both IBH procedures are very
close and much better than that of BH95. For $\rho= 0.8$ STS has
the highest power, followed closely by the oracle, both IBH and
BKY, with a large gap to BH95. Again, one should bear in mind that
STS has the largest number of discoveries $R$, at the cost of
violating strongly the bound of 0.05 on the FDR. Interestingly,
there is no simple monotonicity relationship between the values of
the FDR, $E(V/R^+)$, and the True Discovery Rate $E(S/m_1)$.

Figure \ref{fig:comparison_EVR_and_ESm}e shows
the standard deviation (st.d.) of $V/R^+$ for the independent case, and
Figure \ref{fig:comparison_EVR_and_ESm}f for the
positively dependent case. As can be seen when the $p$-values are
independent, the st.d. is very similar for all the procedures, but
increases steeply as $m_0/m\rightarrow1$. In the case of dependent
$p$-values, the situation becomes worse; for nearly the entire range of
$m_0/m$ the coefficient of variance $\mathit{cv} = \mathit{st.d.}(V/R^+)/E(V/R^+)$ is
greater than 1. Also, as will be mentioned below, for real data the
st.d. of the STS procedure is significantly higher than that of the
IBH. These high values of st.d. result from the FDR definition,
since the expectation of $V/R^+$ takes into account many
realizations with $R=0$ that give, by definition, $V/R^+=0$, making
the distribution of $V/R^+$ very nonsymmetric. A~comparison similar to the
one presented in Figure \ref{fig:comparison_EVR_and_ESm}
for $q=0.05$ is presented in \citet{ZZDsupp2010}, Supplement E, Figure
S4 for $q=0.2$, and provides similar observations. We thus conclude
that for
$\rho=0$ our IBH procedures provide an expected improvement over the
BH95 in
terms of power and saturation of the bound and their performance is comparable
to that of the other adaptive methods tested. For dependent variables STS
violates the bound on $E(V/R^+)$ much more than the IBHlog and the
IBHsum which
violate it only
slightly.

\begin{figure}

\includegraphics{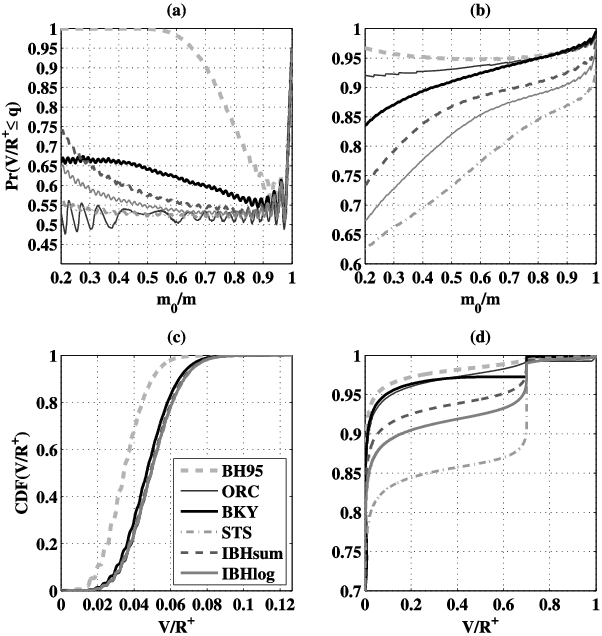}

\caption{\textup{(a)} and \textup{(b)} shows the probability that a single instance
satisfies the
desired FDR level $q$ as a function of $m_0/m$. Results are shown for simulated
data with $m=1000$ hypotheses, the mean of the distribution $P_1$ was
$\mu_1 =
3.5$, and the FDR bound was set to $q=0.05$. Five methods are
compared: ORC, BKY, STS and our two IBH procedures (in the step-down manner).
\textup{(a)} $\rho=0$ and \textup{(b)} $\rho=0.8$. The oscillatory behavior of
some bounds is caused by finite size effects. \textup{(c)} and \textup{(d)} shows the cumulative
distribution function of $V/R^+$ for $m_0/m=0.7$, \textup{(c)} $\rho=0$ and \textup{(d)}
$\rho=0.8$ (obtained from $10^6$ realizations).}\label{fig:m_500_q5_mu1_3p5_Pr_VR_leq_q_comparison_tex}
\end{figure}

\subsection{\texorpdfstring{Applicability for a particular realization.}{Applicability for a particular realization}}
Controlling the FDR at a~level $q$ means that the \textit{average}
fraction of false rejections is no larger than~$q$. It could still
be the case that on average the fraction of false rejections is
controlled, yet
for a large percentage of the realizations one gets many false
rejections and a
high proportion of false discoveries. In contrast to the average behavior
captured by the FDR definition, questions involving the distribution of false
rejections, affecting the behavior of a particular realization, were not
studied much in the literature [a~notable exception is \citet{owenFDvariance}
who studied the variance of $R$]. We therefore set out to address the
issue of
validity of the bound for a particular realization, by calculating for the
synthetic data the probability
$\Pr(\frac{V}{R^+} \leq q)$. This was done for $q=0.05$ for the six
procedures (ORC, BH95, BKY, STS, IBHsum and IBHlog, the latter two in step-down
mode). The probability $\Pr(\frac{V}{R^+} \leq q)$ was estimated by computing,
for each procedure, the fraction of realizations in which we indeed got
$\frac{V}{R^+} \leq q$.
In such a comparison one should bear in mind that a conservative
procedure, such as BH95, restricts the discoveries much more than a
procedure that produces tight bounds (such as the oracle). For
example, looking at Figure \ref{fig:comparison_EVR_and_ESm}a, we see that
the mean value $E(V/R^+)$ of BH95 is much lower than $q=0.05$, and,
hence, the weight of the tail of the distribution of $V/R^+$ values
that ``leaks'' to $V/R^+ > 0.05$ is very small, whereas for the
oracle, which has $E(V/R^+) \approx0.05$, the probability of
exceeding $0.05$ is close to $0.5$, and if we want to guarantee that
$\Pr(V/R^+<B)\approx1$, we must set $B$ at a value which is
significantly larger than the FDR bound $q$. As seen in Figure
\ref{fig:m_500_q5_mu1_3p5_Pr_VR_leq_q_comparison_tex}a, the results
of IBH are slightly more conservative than the oracle in the case of
independence, while all improved procedures have fairly similar
results. In the case of strong dependency, Figure
\ref{fig:m_500_q5_mu1_3p5_Pr_VR_leq_q_comparison_tex}b, the
differences between the procedures are more pronounced; the STS is
the most permissive procedure.

It is very interesting to see that in the case of positive dependent
statistics the probability to violate the bound is smaller, although $E[V/R^+]$
is larger. This is again due to the fact that in these cases we get
$R=0$ for
many realizations, which means that $V/R^+=0$, that is, the variance of
$V/R^+$ is
increased for positive correlations, whereas for the independent case $V/R^+$
is very likely to be close to its expectation. Further study on the distribution
of $V/R^+$ is required in order to shed light on the behavior of different
procedures for particular realizations.
Figure \ref{fig:m_500_q5_mu1_3p5_Pr_VR_leq_q_comparison_tex}c and d
present the
cumulative distribution function (CDF) of $V/R^+$ for a specific set of
parameters, $m=1000,m_0/m=0.7,\mu_1=3.5,q=0.05$, and the different procedures
to be compared, for the independent case (Figure
\ref{fig:m_500_q5_mu1_3p5_Pr_VR_leq_q_comparison_tex}c) and for the
positive dependence case (Figure
\ref{fig:m_500_q5_mu1_3p5_Pr_VR_leq_q_comparison_tex}d).
We would like to emphasize two points: (1) The CDFs of our improved procedures
have very similar behavior to the other improved procedures. (2) While in the
independent case the distribution is close to symmetric, under
dependency the
distribution is very nonsymmetric, and, hence, controlling the mean (of $V/R^+$)
is almost irrelevant.

\section{\texorpdfstring{Application to gene expression data.}{Application to gene expression data}}\label{sec:application}
As an ultimate test for their utility,
we wanted to asses the performance of our new procedures on real life data,
which typically provide complex and unexpected dependency structures
which are
hard to capture in simulations. We therefore applied our procedures that
were described in Section \ref{sec:proposed_procedure} to publicly
available expression data. First, we present in full detail how
our procedures were applied to two data sets. Next, our procedures were
applied to 33 data sets and results were compared with those
obtained by several other procedures: the original BH95 and the
improved bounds of BKY [\citet{BKY}] and STS [\citet{storey2004}] with
$\lambda=0.5$.

\subsection{\texorpdfstring{Detailed application of our procedures.}{Detailed application of our procedures}}
The first data set used is that of \citet{andersson2007} who studied
several types of childhood leukemia. We focus here on the search for
genes whose expression separated 6 patients with normal bone marrow
from 11 T-Cell Acute Lymphoblastic Leukemia patients, which yielded
a large number of discoveries (differentiating genes). The number of
hypotheses (e.g., potentially differentiating probe sets) was
$m=21\mbox{,}288$; the corresponding reported $p$-values were ordered and
plotted on Figure \ref{fig:example_pval}a. Our estimators for $m_0$,
obtained using equations (\ref{eq:Chatm0}) and (\ref{def:mgal}) for this
data, were $\hatm= 7093,\galm=6380$, and the estimated numbers of
discoveries were $m-\hatm\approx14\mbox{,}000, m-\galm\approx15\mbox{,}000$.

The second study, of \citet{pawitan2005} on breast cancer, had a
relatively small number of discoveries. The aim was to find genes
that differentiated early discovery breast cancer cases of poor and
good outcomes, that is, were differentially expressed between tumors obtained
from $38$ subjects that died of the disease and from $121$
patients who were alive. The number of hypotheses was $m=44\mbox{,}611$, and
our $p$-values based estimators for $m_0$ (plotted in Figure
\ref{fig:example_pval}b) were $\hatm=
38\mbox{,}587,\galm=37\mbox{,}580$.

For both studies we have set the desired FDR value at $q=0.1$. We plot in
Figure~\ref{fig:example_pval} the sorted $p$-values $p_{(i)}$ versus
$i/m$ for these two data sets. In each of the two figures we show
three FDR lines; the $\alpha_i$ of BH95 [see equation~(\ref{eq:alpha})]
and the values of $\gamma_i$ corresponding to our two procedures [see
equation~(\ref{def:gamma_i})].

For the first data set the BH95 procedure yields at $q=0.1$ a
large number of $R=0.6065 \times21\mbox{,}288 = 12\mbox{,}912$ discoveries (see
Figure \ref{fig:example_pval}a). When we apply our procedure we
get, at the same FDR, $R' =0.746 \cdot21\mbox{,}288 =15\mbox{,}884$ (for the
\mbox{IBHsum}) discoveries, that is, 23\% more.

The BH95 procedure yields for the second data set (at $q=0.1$)
$R=499$ discoveries. When we apply our procedure we get, at the
same FDR, $R' = 621$ (for the IBHsum) discoveries, that is, 24\% more.

\subsection{\texorpdfstring{Applying our procedures to many data sets.}{Applying our procedures to many data sets}}
\begin{figure}

\includegraphics{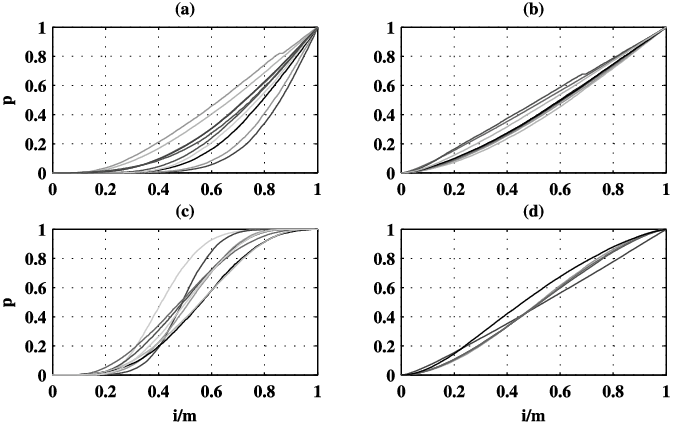}

\caption{Sorted $p$-value vectors from 33 expression data sets of
various cancer-related comparisons: \textup{(a)} two-tailed tests
with large numbers of discoveries, \textup{(b)} two-tailed tests with
small numbers of discoveries, \textup{(c)} one-tailed tests with large numbers of
discoveries, \textup{(d)} one-tailed tests with small numbers of
discoveries.
\label{fig:pval_vectors}}
\end{figure}

We downloaded
from the ONCOMINE website [\citet{rhodes2007}] $p$-value vectors that
were obtained from $33$ comparisons, performed on expression data
from 19 studies of various types of cancer:
\citet{andersson2007}; \citet{basso2005}; \citet{bittner2005};
\citet{bullinger2004}; \citet{choi2007}; \citet{chowdary2006};
\citet{graudens2006}; \citet{koinuma2006}; \citet{laiho2007};
\citet{miller2005}; \citet{pawitan2005}; \citet{ross2003};
\citet{valk2004}; \citet{vandevijver2002}; \citet{wang2005};
\citet{watanabe2006}; \citet{yeoh2002}; \citet{zhao2004}; \citet{zou2002}.
Depending on the biological question at hand, either one or
two-tailed tests are appropriate. Therefore, we applied our
procedures to both test types. We focused on two opposing scenarios:
those with a small number (less than 2\% of $m$, for the
BH95 procedure with $q=0.05$) of discoveries, and those with a large
number (more than 10\% of $m$). The 33 sorted sets of $p_i$ values
are plotted, ver\-sus~$i/m$, in Figure \ref{fig:pval_vectors},
separately for the
four types of comparisons that were made (one/two-tailed test, low/high number
of discoveries).

%

As can be seen in Figure \ref{fig:pval_vectors}, for each type of
comparison the sorted $p$-value curve has a typical shape. In the case
of a large number of discoveries, Figure~\ref{fig:pval_vectors}a
and c, the curve is more convex (and flatter near zero) than in the
case of a small number of discoveries, Figure
\ref{fig:pval_vectors}b and d. Another clear difference is between
the two-tailed (Figure \ref{fig:pval_vectors}a and b) and the
one-tailed (Figure \ref{fig:pval_vectors}c and~d) sorted $p$-value
curves. In the case of two-tailed tests, the entire curve is convex,
while for one-tailed tests the right side of the curve is concave;
the reason is that in the latter case there are very often some hypotheses
that are shifted, with respect to the null hypothesis, in the direction opposite
to the one tested for by the one-sided test (for example, if one looks
for up-regulated genes, there are typically also many down-regulated
genes, which produce very high $p$-values). For detailed treatment of
FDR estimation in the case of one tailed tests see
\citet{poundscheng2006}.\looseness=-1

We compare here the performance of five procedures: the
BH95, BKY, STS, IBHsum and IBHlog (both IBH in the step-down mode).
For each of the improved procedures we determined the ratio between
the number of rejected hypotheses it yielded and the number of
hypotheses rejected by BH95. We present in Table
\ref{table:comparison_real_data} the mean value of this figure of
merit and its standard deviation, calculated for the data sets of
each of the types of comparisons mentioned above, at $q=0.05$ and
$q=0.1$.

\begin{table}
\caption{Comparison of the improvement in power
(ratio between numbers of rejected hypotheses with respect to the
BH95 procedure: $R / R_{BH95}$) of several methods: BKY [\protect\citet{BKY}], STS
[\protect\citet{storey2004}], IBHsum and IBHlog in the step-down~version. Mean
values and standard deviations (in parentheses) are given for~each~of~the~four types of comparisons}\label{table:comparison_real_data}
\begin{tabular*}{\tablewidth}{@{\extracolsep{\fill}}lcccc@{}}
\hline
$\bolds{q}$ & \textbf{BKY} & \textbf{STS} & \textbf{IBHsum} & \textbf{IBHlog} \\
\hline
\multicolumn{5}{@{}l@{}}{(a) Two-tailed, large number of discoveries (10 studies)} \\
0.05 & 1.110 & 1.239 & 1.200 & 1.222 \\
& (0.043) & (0.138) & (0.110) & (0.130) \\
0.1 & 1.155 & 1.258 & 1.213 & 1.237 \\
& (0.057) & (0.117) & (0.087) & (0.102) \\[3pt]
\multicolumn{5}{@{}l@{}}{(b) Two-tailed, small number of discoveries (10 studies)} \\
0.05 & 1.003 & 1.316 & 1.231 & 1.291 \\
& (0.003) & (0.197) & (0.140) & (0.179) \\
0.1 & 1.017 & 1.308 & 1.230 & 1.275 \\
& (0.027) & (0.161) & (0.117) & (0.137) \\[3pt]
\multicolumn{5}{@{}l@{}}{(c) One-tailed, large number of discoveries (8 studies)} \\
0.05 & 1.049 & 1.011 & 1.014 & 0.108 \\
& (0.019) & (0.033) & (0.026) & (0.306) \\
0.1 & 1.062 & 1.012 & 1.014 & 0.108 \\
& (0.026) & (0.0340 & (0.024) & (0.305) \\[3pt]
\multicolumn{5}{@{}l@{}}{(d) One-tailed, small number of discoveries (5 studies)} \\
0.05 & 0.998 & 1.027 & 1.025 & 0.882 \\
& (0.020) & (0.052) & (0.017) & (0.123) \\
0.1 & 1.004 & 1.028 & 1.031 & 0.888 \\
& (0.031) & (0.079) & (0.022) & (0.120) \\
\hline
\end{tabular*}
\end{table}

Inspection of Table \ref{table:comparison_real_data} reveals that
for types (a), (b)---of two-tailed tests, irrespective of the number of
discoveries and FDR level, STS and both IBH procedures give
significantly higher improvement over BH95 than the BKY procedure,
with STS performing slightly better than IBHlog, followed by
IBHsum. For the one-tailed test with large numbers of discoveries
[type (c)] the mean improvement of BKY is the highest, while STS and
IBHsum are quite similar. IBHlog fails dramatically in this case
due to the abundance of $p$-values close to one, giving an over-estimation
of $m_0$. For type (d), one-tailed tests with a small number of discoveries,
IBHsum is slightly better than STS and both yield a significantly higher
improvement than BKY. In all four types
and for all values of FDR, the standard deviations of $V/R^+$
of the STS method are significantly higher than those of
BKY and the IBHsum procedures. Furthermore, as shown in Section
\ref{subsec:comparison_of_methods} (see Figure
\ref{fig:comparison_EVR_and_ESm}b), in the
case of positively dependent test statistics the STS procedure
loses control of the FDR in a much more drastic manner than our
IBH procedures. Since we expect that correlations between the
expression profiles of different genes will be present in most
data, the STS method may produce unreliable values of the figure
of merit presented here.

In summary, our IBH procedures constitute in all cases a
significant improvement over the original BH95; in all but one of
the comparison types the improvement is significantly better than
that of the BKY method. Comparison with STS yields mixed results,
but the edge of STS over IBH in two of the four comparison types
is overshadowed by the fact that STS does not provide a reliable
bound for data sets with positive correlations between probe sets,
while IBH remains reliable.

\section{\texorpdfstring{Discussion.}{Discussion}}
We addressed the problem of controlling the False Discovery Rate
in the case of a large number of comparisons, or hypotheses to be
tested simultaneously. Providing a reliable and possibly tight
bound on the FDR is an issue of major importance for analysis of
high-throughput biological data, such as obtained using gene
expression microarrays. We presented here two estimators of $m_0$, the
number of
true null hypotheses. We proved that both estimators can be used for FDR
estimation and, more importantly, for FDR control. Thus, we added two procedures
to the rather limited repertoire of improved FDR procedures for which
control of
the FDR is known to hold. Our proof of control relies on a general theorem,
which provides a bound on the FDR for improved procedure using any estimator
$\hatm(p_1,\ldots,p_m)$ provided a condition of monotonicity is satisfied,
and one
is able to bound the reciprocal mean of the estimator. In addition, we
proved a
novel result, that FDR procedures satisfy a monotonicity property under some
very plausible assumptions. As a corollary of this theorem, we show
that any
bound on the FDR that was proved for the step-up procedure, holds also
for the
more conservative step-down procedure as well. Our
proofs of control hold only for the independent case. For the dependent case,
results for control are even more scarce, and limited to certain
specific types
of dependency. We therefore studied the behavior of our procedures,
compared to
others known from the literature, under dependency, using simulations.
In addition to studying behavior under dependency, our simulations also enabled
us to understand the distribution of the fraction of false hypotheses,
and, in particular, the probability of violating the bound for a particular
given realization. Further research on this aspect of comparing procedures
is needed and we expect it to provide interesting new insights and
measures for
comparisons of different procedures.
We finally applied our procedures, as well as several others, to a
large number
of cancer-related expression data sets. For both real and simulated data,
our new procedures provided more rejections (separating genes) than the similar
list of Benjamini and Hochberg and the very recently introduced
improved bound
of BKY [\citet{BKY}], for a fixed desired value of the FDR. In some
cases the
improved bound of STS [\citet{storey2004}] gives more rejection than
our method,
but as we have shown on synthetic data, when there are positive correlations,
STS loses control of the FDR in a much more pronounced way than our procedure.
To summarize: a researcher may either obtain a desired number of differentially
expressed genes at a lower FDR, or get a longer list of such genes at the
desired FDR level, at no added computational cost. We recommend using
our IBHlog
procedure for two-tailed tests, and IBHsum procedure for a one-tailed
test, to
increase discovery power while controlling FDR levels.

\section*{\texorpdfstring{Acknowledgments.}{Acknowledgments}}
We thank Y. Benjamini for most helpful discussions and
encouragement, and A. Gubichev for help with programming.

\begin{supplement}
\stitle{Supplementary material for: FDR control with adaptive
procedures and FDR
monotonicity}
\slink[doi]{10.1214/10-AOAS399SUPP}
\slink[url]{http://lib.stat.cmu.edu/aoas/399/supplement.pdf}
\sdatatype{.pdf}
\sdescription{In this supplementary file we provide proofs of
the claims and theorem presented in the paper, together with technical details
regarding the proposed estimator and of the simulations performed. The document
includes the following sections:
Supplement A: Proof of Theorem 2.3.
Supplement B: Designing the IBHsum estimator.
Supplement C: Proof of Claim 3.1.
Supplement D: Proof of the monotonicity theorem.
Supplement E: Details of the simulations.}
\end{supplement}


\printaddresses


\begin{thebibliography}{99}

\bibitem[\protect\citeauthoryear{Andersson {et~al.}}{2007}]{andersson2007}
\textsc{Andersson}, A., \textsc{Ritz}, C., \textsc{Lindgren}, D.,
\textsc{Ed{\'{e}}n}, P., \textsc{Lassen}, C., \textsc{Heldrup}, J.,
\textsc{Olofsson},~T., \textsc{R\r{a}de}, J., \textsc{Fontes}, M.,
\textsc{Porwit-Macdonald}, A., \textsc{Behrendtz}, M.,
\textsc{H{\"{o}}glund},~M., \textsc{Johansson}, B. and \textsc
{Fioretos}, T. (2007).
Microarray-based classification of a consecutive series of 121
childhood acute leukemias: Prediction of leukemic and genetic subtype as well as
of minimal residual disease status.
\textit{Leukemia} \textbf{21} 1198--1203.

\bibitem[\protect\citeauthoryear{Aven and Jensen}{1999}]{Aven01}
\textsc{Aven}, T. and \textsc{Jensen}, U. (1999).
\textit{Stochastic Models in Reliability}.
Springer, New York.
\MR{1679540}

\bibitem[\protect\citeauthoryear{Basso {et~al.}}{2005}]{basso2005}
\textsc{Basso}, K., \textsc{Margolin}, A.~A., \textsc{Stolovitzky},
G., \textsc{Klein}, U., \textsc{Dalla-Favera}, R. and
\textsc{Califano}, A. (2005). Reverse engineering of regulatory networks in human {B}
cells. \textit{Nat. Genet.} \textbf{37} 382--390.

\bibitem[\protect\citeauthoryear{Benjamini and Hochberg}{1995}]{BH95}
\textsc{Benjamini}, Y. and \textsc{Hochberg}, Y. (1995). Controlling the false discovery rate: A practical and powerful
              approach to multiple testing.
\textit{J. Roy. Statist. Soc. Ser. B} \textbf{57} 289--300.
\MR{1325392}

\bibitem[\protect\citeauthoryear{Benjamini, Krieger and Yekutieli}{2006}]{BKY}
\textsc{Benjamini}, Y., \textsc{Krieger}, A.~M. and \textsc{Yekutieli}, D.
(2006). Adaptive linear step-up procedures that control the false
              discovery rate.
\textit{Biometrica} \textbf{93} 491--507.
\MR{2261438}

\bibitem[\protect\citeauthoryear{Benjamini and Yekutieli}{2001}]{BY2001}
\textsc{Benjamini}, Y. and \textsc{Yekutieli}, D. (2001). The control of the false discovery rate in multiple testing
              under dependency.
\textit{Ann. Statist.} \textbf{29} 1165--1168.
\MR{1869245}

\bibitem[\protect\citeauthoryear{Bittner}{2005}]{bittner2005}
\textsc{Bittner}, M. (2005). {A} window on the dynamics of biological
switches.
\textit{Nat. Biotechnol.} \textbf{23} 183--184.

\bibitem[\protect\citeauthoryear{Bullinger {et~al.}}{2004}]{bullinger2004}
\textsc{Bullinger}, L., \textsc{D{\"{o}}hner}, K., \textsc{Bair},
E., \textsc{Fr{\"{o}}hling}, S., \textsc{Schlenk}, R.~F.,
\textsc{Tibshirani}, R., \textsc{D{\"{o}}hner}, H. and \textsc
{Pollack}, J.~R. (2004). {Use} of gene-expression profiling to identify prognostic subclasses in adult acute myeloid
leukemia. \textit{N. Engl. J. Med.} \textbf{350} 1605--1616.

\bibitem[\protect\citeauthoryear{Choi {et~al.}}{2007}]{choi2007}
\textsc{Choi}, Y.~L., \textsc{Tsukasaki}, K., \textsc{O'Neill},
M.~C., \textsc{Yamada}, Y., \textsc{Onimaru}, Y., \textsc{Matsumoto},~K.,
\textsc{Ohashi}, J., \textsc{Yamashita}, Y., \textsc{Tsutsumi},
S., \textsc{Kaneda}, R., \textsc{Takada}, S.,
\textsc{Aburatani},~H., \textsc{Kamihira}, S., \textsc{Nakamura},
T., \textsc{Tomonaga}, M. and \textsc{Mano}, H. (2007). {A} genomic analysis of adult {T-cell}
leukemia. \textit{Oncogene} \textbf{26} 1245--1255.

\bibitem[\protect\citeauthoryear{Chowdary {et~al.}}{2006}]{chowdary2006}
\textsc{Chowdary}, D., \textsc{Lathrop}, J., \textsc{Skelton}, J.,
\textsc{Curtin}, K., \textsc{Briggs}, T., \textsc{Zhang}, Y.,
\textsc{Yu},
J., \textsc{Wang},~Y. and \textsc{Mazumder}, A. (2006).
{Prognostic} gene expression signatures can be measured in tissues collected in {RNA later}
preservative. \textit{J. Mol. Diagn.} \textbf{8} 31--39.

\bibitem[\protect\citeauthoryear{Gavrilov, Benjamini and Sarkar}{2009}]{gavrilov2009}
\textsc{Gavrilov}, Y., \textsc{Benjamini}, Y. and \textsc{Sarkar},
S.~K. (2009). An adaptive step-down procedure with proven {FDR} control
              under independence.
\textit{Ann. Statist.} \textbf{37} 619--629.
\MR{2502645}

\bibitem[\protect\citeauthoryear{Genovese and
Wasserman}{2002}]{Genovese2002}
\textsc{Genovese}, C. and \textsc{Wasserman}, L. (2002). Operating characteristics and extensions of the false
              discovery rate procedure.
\textit{J. R. Stat. Soc. Ser. B Stat. Methodol.} \textbf{64} 499--517.
\MR{1924303}

\bibitem[\protect\citeauthoryear{Graudens {et~al.}}{2006}]{graudens2006}
\textsc{Graudens}, E., \textsc{Boulanger}, V., \textsc{Mollard}, C.,
\textsc{Mariage-Samson}, R., \textsc{Barlet}, X.,
\textsc{Gr{\'{e}}my},~G., \textsc{Couillault}, C., \textsc{Laj{\'
{e}}mi}, M., \textsc{Piatier-Tonneau}, D.,
\textsc{Zaborski}, P., \textsc{Eveno},~E., \textsc{Auffray}, C. and
\textsc{Imbeaud}, S. (2006). {Deciphering} cellular states of innate tumor drug
responses. \textit{Genome Biol.} \textbf{7} R19--R19.

\bibitem[\protect\citeauthoryear{Koinuma {et~al.}}{2006}]{koinuma2006}
\textsc{Koinuma}, K., \textsc{Yamashita}, Y., \textsc{Liu}, W.,
\textsc{Hatanaka}, H., \textsc{Kurashina}, K., \textsc{Wada}, T.,
\textsc{Takada},~S., \textsc{Kaneda}, R., \textsc{Choi}, Y.~L.,
\textsc{Fujiwara}, S.-I., \textsc{Miyakura}, Y., \textsc{Nagai},
H. and \textsc{Mano}, H. (2006). {Epigenetic} silencing of {AXIN2} in colorectal carcinoma with microsatellite
instability. \textit{Oncogene} \textbf{25} 139--146.

\bibitem[\protect\citeauthoryear{Laiho {et~al.}}{2007}]{laiho2007}
\textsc{Laiho}, P., \textsc{Kokko}, A., \textsc{Vanharanta}, S.,
\textsc{Salovaara}, R., \textsc{Sammalkorpi}, H.,
\textsc{J{\"{a}}rvinen},~H., \textsc{Mecklin}, J.-P., \textsc
{Karttunen}, T.~J., \textsc{Tuppurainen}, K.,
\textsc{Davalos}, V., \textsc{Schwartz}, S., \textsc{Arango}, D.,
\textsc{M{\"{a}}kinen}, M.~J. and \textsc{Aaltonen},
L.~A. (2007). {Serrated} carcinomas form a subclass of colorectal cancer with distinct molecular
basis.
\textit{Oncogene} \textbf{26} 312--320.

\bibitem[\protect\citeauthoryear{Miller {et~al.}}{2005}]{miller2005}
\textsc{Miller}, L.~D., \textsc{Smeds}, J., \textsc{George}, J.,
\textsc{Vega}, V.~B., \textsc{Vergara}, L., \textsc{Ploner}, A.,
\textsc{Pawitan},~Y., \textsc{Hall}, P., \textsc{Klaar}, S., \textsc
{Liu}, E.~T. and \textsc{Bergh}, J. (2005).
An expression signature for p53 status in human breast cancer
predicts mutation status, transcriptional effects, and patient
survival.
\textit{Proc. Natl. Acad. Sci. USA} \textbf{102} 13550--13555.

\bibitem[\protect\citeauthoryear{Owen}{2005}]{owenFDvariance}
\textsc{Owen}, A.~B. (2005). Variance of the number of false
discoveries.
\textit{J. Roy. Statist. Soc. Ser. B} \textbf{67}
411--426.
\MR{2155346}

\bibitem[\protect\citeauthoryear{Pawitan {et~al.}}{2005}]{pawitan2005}
\textsc{Pawitan}, Y., \textsc{Bj{\"{o}}hle}, J., \textsc{Amler}, L.,
\textsc{Borg}, A.-L., \textsc{Egyhazi}, S., \textsc{Hall}, P.,
\textsc{Han}, X., \textsc{Holmberg}, L., \textsc{Huang}, F., \textsc
{Klaar}, S., \textsc{Liu}, E.~T., \textsc{Miller}, L.,
\textsc{Nordgren}, H., \textsc{Ploner}, A., \textsc{Sandelin}, K.,
\textsc{Shaw}, P.~M., \textsc{Smeds}, J., \textsc{Skoog}, L.,
\textsc{Wedr{\'{e}}n}, S. and \textsc{Bergh}, J. (2005).
{Gene} expression profiling spares early breast cancer patients from
adjuvant therapy: Derived and validated in two population-based
cohorts. \textit{Breast Cancer Res.} \textbf{7} R953--R964.

\bibitem[\protect\citeauthoryear{Pounds and Cheng}{2006}]{poundscheng2006}
\textsc{Pounds}, S. and \textsc{Cheng}, C. (2006).
{Robust} estimation of the false discovery rate.
\textit{Bioinformatics} \textbf{22} 1979--1987.

\bibitem[\protect\citeauthoryear{Reiner}{2007}]{Reiner2007}
\textsc{Reiner}, A. (2007). FDR control by the {BH} procedure for two-sided correlated
              tests with implications to gene expression data analysis.
\textit{Biom. J.} \textbf{49} 107--126.
\MR{2339220}

\bibitem[\protect\citeauthoryear{Rhodes {et~al.}}{2007}]{rhodes2007}
\textsc{Rhodes}, D.~R., \textsc{Kalyana-Sundaram}, S., \textsc
{Mahavisno}, V., \textsc{Varambally}, R., \textsc{Yu}, J.,
\textsc{Briggs}, B.~B., \textsc{Barrette}, T.~R., \textsc{Anstet},
M.~J., \textsc{Kincead-Beal}, C., \textsc{Kulkarni},
P., \textsc{Varambally},~S., \textsc{Ghosh}, D. and \textsc
{Chinnaiyan}, A.~M. (2007).
{Oncomine} 3.0: Genes, pathways, and networks in a collection of 18,000 cancer gene expression
profiles.
\textit{Neoplasia} \textbf{9} 166--180.

\bibitem[\protect\citeauthoryear{Ross {et~al.}}{2003}]{ross2003}
\textsc{Ross}, M.~E., \textsc{Zhou}, X., \textsc{Song}, G., \textsc
{Shurtleff}, S.~A., \textsc{Girtman}, K., \textsc{Williams},
W.~K., \textsc{Liu}, H.-C., \textsc{Mahfouz}, R., \textsc{Raimondi},
S.~C., \textsc{Lenny}, N., \textsc{Patel}, A. and
\textsc{Downing},~J.~R. (2003).
{Classification} of pediatric acute lymphoblastic leukemia by gene expression
profiling.
\textit{Blood} \textbf{102} 2951--2959.

\bibitem[\protect\citeauthoryear{Storey}{2002}]{storey2002}
\textsc{Storey}, J.~D. (2002).
{A} direct approach to false discovery rate.
\textit{J. Roy. Statist. Soc. Ser. B} \textbf{64} 479--498.
\MR{1924302}

\bibitem[\protect\citeauthoryear{Storey, Taylor and Siegmund}{2004}]{storey2004}
\textsc{Storey}, J.~D., \textsc{Taylor}, J.~E. and \textsc{Siegmund}, D.
(2004). Strong control, conservative point estimation and simultaneous
              conservative consistency of false discovery rates: A~unified
              approach.
\textit{J. Roy. Statist. Soc. Ser. B} \textbf{66} 187--205.
\MR{2035766}

\bibitem[\protect\citeauthoryear{Valk {et~al.}}{2004}]{valk2004}
\textsc{Valk}, P. J.~M., \textsc{Verhaak}, R. G.~W., \textsc
{Beijen}, M.~A., \textsc{Erpelinck}, C. A.~J.,
\textsc{Barjesteh van Waalwijk~van Doorn-Khosrovani}, S., \textsc
{Boer}, J.~M., \textsc{Beverloo},
H.~B., \textsc{Moorhouse},~M.~J., \textsc{van~der Spek}, P.~J.,
\textsc{L{\"{o}}wenberg}, B. and
\textsc{Delwel}, R. (2004).
{Prognostically} useful gene-expression profiles in acute myeloid
leukemia. \textit{N. Engl. J. Med.} \textbf{350} 1617--1628.

\bibitem[\protect\citeauthoryear{van~de Vijver
{et~al.}}{2002}]{vandevijver2002}
\textsc{van~de Vijver}, M.~J., \textsc{He}, Y.~D., \textsc{van't
Veer}, L.~J., \textsc{Dai}, H., \textsc{Hart}, A. A.~M.,
\textsc{Voskuil},~D.~W., \textsc{Schreiber}, G.~J., \textsc
{Peterse}, J.~L., \textsc{Roberts}, C., \textsc{Marton}, M.~J.,
\textsc{Parrish}, M., \textsc{Atsma}, D., \textsc{Witteveen}, A.,
\textsc{Glas}, A., \textsc{Delahaye}, L., \textsc{van~der Velde},~%
T., \textsc{Bartelink},~H., \textsc{Rodenhuis}, S., \textsc
{Rutgers}, E.~T., \textsc{Friend}, S.~H. and
\textsc{Bernards}, R. (2002).
{A} gene-expression signature as a predictor of survival in breast
cancer. \textit{N. Engl. J. Med.} \textbf{347} 1999--2009.

\bibitem[\protect\citeauthoryear{Wang {et~al.}}{2005}]{wang2005}
\textsc{Wang}, Y., \textsc{Klijn}, J. G.~M., \textsc{Zhang}, Y.,
\textsc{Sieuwerts}, A.~M., \textsc{Look}, M.~P., \textsc{Yang}, F.,
\textsc{Talantov},~D., \textsc{Timmermans}, M., \textsc{Meijer-van
Gelder}, M.~E., \textsc{Yu}, J., \textsc{Jatkoe}, T.,
\textsc{Berns}, E. M. J.~J., \textsc{Atkins}, D. and \textsc
{Foekens}, J.~A. (2005).
{Gene-expression} profiles to predict distant metastasis of lymph-node-negative primary breast
cancer. \textit{Lancet} \textbf{365} 671--679.

\bibitem[\protect\citeauthoryear{Watanabe {et~al.}}{2006}]{watanabe2006}
\textsc{Watanabe}, T., \textsc{Kobunai}, T., \textsc{Toda}, E.,
\textsc{Yamamoto}, Y., \textsc{Kanazawa}, T., \textsc{Kazama},~Y.,
\textsc{Tanaka},~J., \textsc{Tanaka},~T., \textsc{Konishi}, T.,
\textsc{Okayama}, Y., \textsc{Sugimoto}, Y., \textsc{Oka}, T.,
\textsc{Sasaki},~S., \textsc{Muto}, T. and \textsc{Nagawa}, H. (2006).
{Distal} colorectal cancers with microsatellite instability {(MSI)}
display distinct gene expression profiles that are different from proximal
{MSI} cancers. \textit{Cancer Res.} \textbf{66} 9804--9808.

\bibitem[\protect\citeauthoryear{Yekutieli and
Benjamini}{1999}]{yekutieli1999}
\textsc{Yekutieli}, D. and \textsc{Benjamini}, Y. (1999). Resampling-based false discovery rate controlling multiple
              test procedures for correlated test statistics.
\textit{J. Statist. Plann. Inference} \textbf{82} 171--196.
\MR{1736442}

\bibitem[\protect\citeauthoryear{Yeoh {et~al.}}{2002}]{yeoh2002}
\textsc{Yeoh}, E.-J., \textsc{Ross}, M.~E., \textsc{Shurtleff},
S.~A., \textsc{Williams}, W.~K., \textsc{Patel}, D.,
\textsc{Mahfouz}, R., \textsc{Behm}, F.~G., \textsc{Raimondi},
S.~C., \textsc{Relling}, M.~V., \textsc{Patel}, A., \textsc{Cheng},
C., \textsc{Campana}, D., \textsc{Wilkins}, D., \textsc{Zhou}, X.,
\textsc{Li}, J., \textsc{Liu}, H., \textsc{Pui}, C.-H., \textsc{Evans},
W.~E., \textsc{Naeve}, C., \textsc{Wong}, L. and \textsc{Downing},
J.~R. (2002).
Classification, subtype discovery, and prediction of outcome in
pediatric acute lymphoblastic leukemia by gene expression profiling.
\textit{Cancer Cell} \textbf{1} 133--143.

\bibitem[\protect\citeauthoryear{Zeisel, Zuk and Domany}{2010}]{ZZDsupp2010}
\textsc{Zeisel}, A., \textsc{Zuk}, O. and \textsc{Domany}, E. (2010).
Supplement to ``FDR control with adaptive procedures and FDR
monotonicity.'' DOI:
\href{http://dx.doi.org/10.1214/10-AOAS399SUPP}{10.1214/10-AOAS399SUPP}.

\bibitem[\protect\citeauthoryear{Zhao {et~al.}}{2004}]{zhao2004}
\textsc{Zhao}, H., \textsc{Langer{\o}d}, A., \textsc{Ji}, Y.,
\textsc{Nowels}, K.~W., \textsc{Nesland}, J.~M., \textsc{Tibshirani},
R., \textsc{Bukholm}, I.~K., \textsc{K\r{a}resen}, R., \textsc
{Botstein}, D., \textsc{B{\o}rresen-Dale}, A.-L.
and \textsc{Jeffrey},~S.~S. (2004).
{Different} gene expression patterns in invasive lobular and ductal carcinomas of the
breast. \textit{Mol. Biol. Cell} \textbf{15} 2523--2536.

\bibitem[\protect\citeauthoryear{Zou {et~al.}}{2002}]{zou2002}
\textsc{Zou}, T.-T., \textsc{Selaru}, F.~M., \textsc{Xu}, Y.,
\textsc{Shustova}, V., \textsc{Yin}, J., \textsc{Mori}, Y., \textsc
{Shibata},
D., \textsc{Sato},~F., \textsc{Wang}, S., \textsc{Olaru}, A.,
\textsc{Deacu}, E., \textsc{Liu}, T.~C., \textsc{Abraham}, J.~M. and
\textsc{Meltzer},~S.~J. (2002).
{Application} of {cDNA} microarrays to generate a molecular taxonomy
capable of distinguishing between colon cancer and normal colon.
\textit{Oncogene} \textbf{21} 4855--4862.

\end{thebibliography}
\end{document}